\documentclass[aps,twocolumn,prc,superscriptaddress,noshowpacs,nofootinbib,noshowkeys,floatfix]{revtex4}
\usepackage[dvips]{graphics,graphicx}
\usepackage[colorlinks=true,linktocpage=true,linkcolor=blue,citecolor=blue]{hyperref}
\usepackage[usenames,dvipsnames]{color}
\usepackage{amsmath, amssymb}
\usepackage{multirow}
\usepackage{longtable}
\usepackage{color}
\usepackage{xcolor}
\usepackage{cleveref}
\usepackage[normalem]{ulem}  

\renewcommand\sout{\bgroup \color{blue} \ULdepth=-.5ex \ULset}

\begin{document}

\title{{\Large Development of a water-based cooling system for the Muon Chamber detector system of the CBM experiment}}

\author{Sumit Kumar Kundu}
\email{phd1701151005@iiti.ac.in}
\affiliation{Department of Physics, School of Basic Sciences, Indian Institute of Technology Indore, Indore 453552, INDIA}

\author{Saikat Biswas}
\affiliation{Department of Physics, Bose Institute, EN-80, Sector-V, Salt Lake, Kolkata, 700091, INDIA}
\author{Subhasis Chattopadhyay}
\affiliation{Variable Energy Cyclotron Centre, 1/AF, Bidhan Nagar, Kolkata, 700064, INDIA}
\affiliation{Homi Bhabha National Institute, Mumbai, INDIA}
\author{Supriya Das}
\affiliation{Department of Physics, Bose Institute, EN-80, Sector-V, Salt Lake, Kolkata, 700091, INDIA}
\author{Anand Kumar Dubey}
\affiliation{Variable Energy Cyclotron Centre, 1/AF, Bidhan Nagar, Kolkata, 700064, INDIA}
\affiliation{Homi Bhabha National Institute, Mumbai, INDIA}
\author{Chandrasekhar Ghosh}
\affiliation{Variable Energy Cyclotron Centre, 1/AF, Bidhan Nagar, Kolkata, 700064, INDIA}
\affiliation{Homi Bhabha National Institute, Mumbai, INDIA}
\author{Ajit Kumar}
\affiliation{Variable Energy Cyclotron Centre, 1/AF, Bidhan Nagar, Kolkata, 700064, INDIA}
\affiliation{Homi Bhabha National Institute, Mumbai, INDIA}
\author{Ankhi Roy}
\affiliation{Department of Physics, School of Basic Sciences, Indian Institute of Technology Indore, Indore 453552, INDIA}
\author{Jogender Saini}
\affiliation{Variable Energy Cyclotron Centre, 1/AF, Bidhan Nagar, Kolkata, 700064, INDIA}
\author {Susnata Seth}
\affiliation{Department of Physics, Bose Institute, EN-80, Sector-V, Salt Lake, Kolkata, 700091, INDIA}
\author{Sidharth Kumar Prasad}
\email{sprasad@jcbose.ac.in}
\affiliation{Department of Physics, Bose Institute, EN-80, Sector-V, Salt Lake, Kolkata, 700091, INDIA}


\begin{abstract}

A water-based cooling system is being investigated to meet the cooling requirement of the Gas Electron Multiplier (GEM) based Muon Chamber (MuCh) detector system of the Compressed Baryonic Matter (CBM) experiment at GSI, Germany. The system is based on circulating cold water through the channels inside an aluminium plate. The aluminium plate is attached to a GEM chamber. A feasibility study is conducted on one small and two real-size prototype cooling plates. A microcontroller based unit has been built and integrated into the system to achieve automatic control and monitoring of temperature on plate surface. The real-size prototypes have been used in a test beam experiment at the CERN SPS (Super Proton Synchrotron) with the lead beam on a lead target. A setup using three prototype modules has been prepared in the lab for testing in a simulated real life environment. This paper discusses the working principle, mechanical design, fabrication, and test results of the cooling prototypes in detail.
\end{abstract}

\maketitle

\section*{I. Introduction}
\label{secIntroduction}
A dimuon detector system, the Muon Chamber (MuCh)~\cite{citeMuchCbmTdr}, will be installed in the CBM experiment~\cite{citeCBMexp} at the Facility for Anti-proton and Ion Research (FAIR)~\cite{citeFAIR} at GSI, Germany, for identification and tracking of dimuons produced from the decay of low mass vector mesons and J/$\Psi$ in high energy heavy-ion collisions in the beam energy range of 2 to 35A GeV. Since the lepton pairs formed in these collisions are sensitive probes to the fireball, measuring them in the CBM experiment has become a crucial aspect of the research program~\cite{citeMuchCbmPhysBook}.
Detection of muons is highly challenging, particularly in the low momentum range and in a high particle density environment. CBM overcomes this challenge by instrumenting hadron absorbers segmented in several layers and installing a triplet of tracking detector planes in the gaps in between each absorber pairs.  
The MuCh detector system (absorbers and detector stations) needs to be very compact to reduce the background muons from the weak decays of mesons (pions and kaons) which implies a minimum gap for tracking stations between the absorbers~\cite{citeMuchCbmTdr}.
Each tracking station has three layers (triplets) of detector chambers divided into trapezoidal sector-shaped modules.
The curvature of a station determines the number of sectors (or modules) per plane, and the particle flux determines the detector technology employed in different stations. 
The first and second stations of MuCh detector system will house Gas Electron Multiplier (GEM)~\cite{citeGem} based detector modules. 
  
The Front End electronic Board (FEB) of the MuCh detector system uses a custom-built self-triggered Application-Specific Integrated Circuit (ASIC) which provides both timing and ADC information of each incoming signal to its channel. The FEBs of one MuCh module (sector) dissipates $\sim$90~W of heat while in operation, which implies on an average $\sim$4.3~kW and $\sim$5.4~kW of heat generation by the first (consisting of 16~$\times$~3~=~48~modules) and the second (composed of 20~$\times$~3~=~60~modules) MuCh stations respectively. In addition, as discussed earlier, to meet the compactness requirement, the MuCh stations are to be placed within small gaps in between the absorbers. This large amount of dissipated heat within a confined space results in rise in temperature at the surface of the FEBs and its vicinity.
The FEBs used for the MuCh detector system are sensitive to the surrounding temperature and have a favorable temperature range of 20$^{\circ}$C -- 25$^{\circ}$C for their operation. 
The detectors for the first two stations of the MuCh detector system are GEM-based, as discussed earlier, and the detector's gain is found to depend on the ambient temperature~\cite{citeGemGainTempDepend,citeGemGainTempDepend_2}.
Therefore, the stable running of the detector and FEBs requires a steady ambient temperature within the acceptable range, which in turn demands the necessity of continuous draining out of the generated heat. To meet the above requirement, a cooling system using demineralized water as a coolant is under investigation. The use of an air-based cooling system is ruled out because the interconnections of the FEBs are made using the wire bonding technique, which has the disadvantage of connections getting loosened or broken due to vibrations because of airflow. Considering the heat load, availability, and cost-effectiveness, demineralized water is preferred over other coolants such as methane, liquid nitrogen, C$\rm {O_2}$, etc.

It is also essential that the cooling system does not bring down the temperature at the surface of the FEB below the dew point which might result in the formation of water droplets from the condensation of vapor which is detrimental to the electronics. An automated mechanism is therefore required to control the coolant flow thereby keeping the temperature around a fixed value. To achieve this, a microcontroller based unit is being developed as an integral part of the cooling system.

In this paper, we report the working principle of the cooling system, mechanical design, fabrication, test performances of the prototypes, and the details of the control unit. This particular system is developed keeping MuCh detector system of the CBM experiment in mind. However the basic principle and technique studied here, can be used elsewhere also with proper customization of various parameters such as shape, size, thickness, and material type of the cooling plate. In general, this type of cooling system can be a good solution for places where large heat is generated in a very confined space.

 The paper is organized as follows. Section~\ref{SecWorkingPrinciple} presents the system's working principle. Validation of proof-of-principle of the conceptual design of the cooling system using a stand-alone small-size copper-based prototype is explained in section~\ref{SecSmallSizePrototype}. Section~\ref{SecMechanicalDesign} discusses in detail the mechanical design and fabrication of produced samples of the cooling plate. The working principle and implementation of the control unit are discussed in section~\ref{SecFlowControl}.
 Section~\ref{SecSetupPerformance} briefs the experimental setup and the test performances for real-size prototypes and section~\ref{SecResult} discusses the setup and result of feasibility study of water distribution with multiple prototypes. Section~\ref{SecSummaryAndOutlook}
summarizes the outcome of this work and discusses the future plan.  
\begin{figure}[h]
  \centering
  \includegraphics[width=9cm]{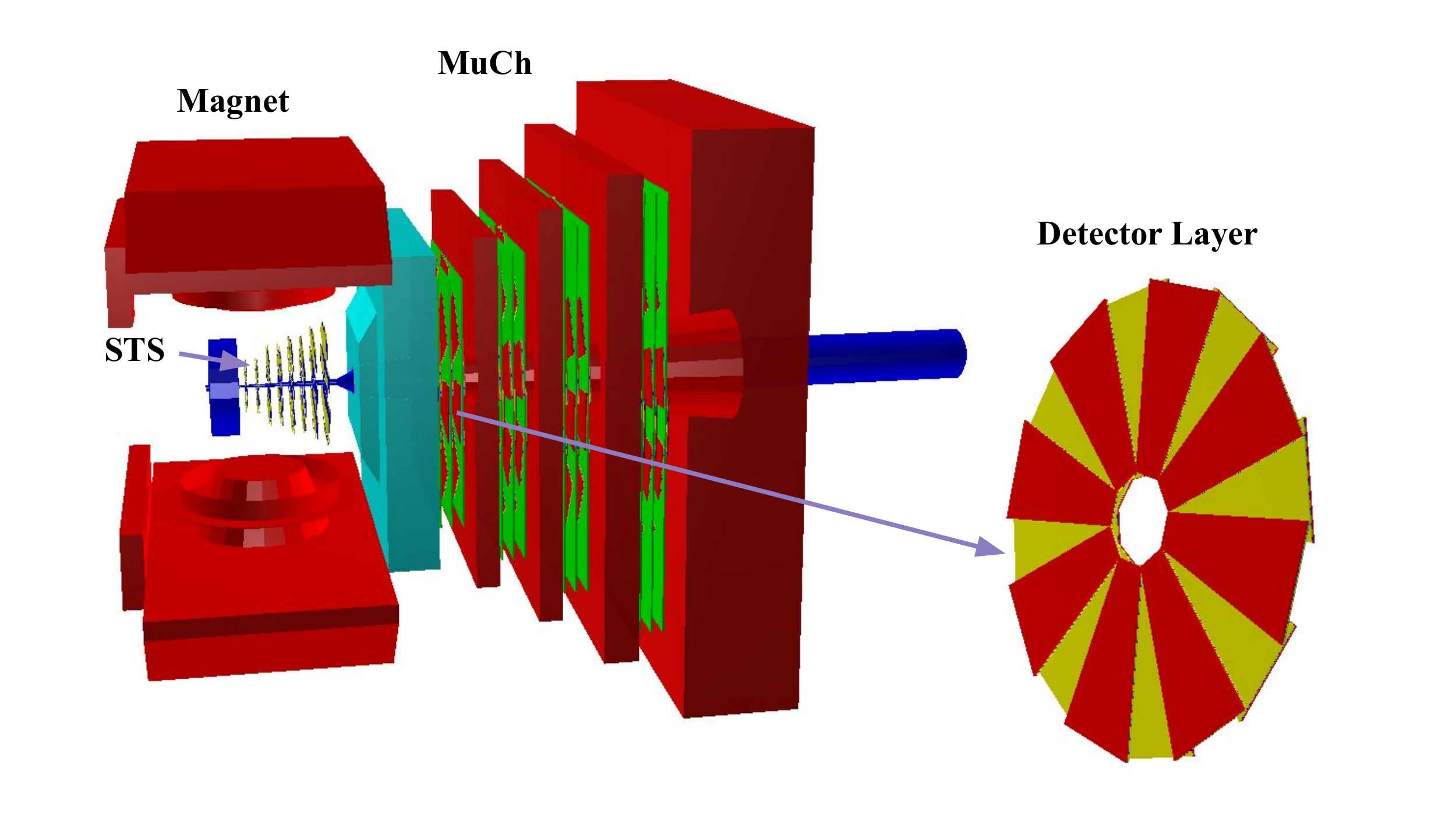}
  \caption{The setup of the CBM experiment along with the MuCh detector system as implemented in the simulation.}
  \label{much_detector_setup}
\end{figure}

\section*{II. Working principle and design considerations}
\label{SecWorkingPrinciple}
As discussed earlier, MuCh detector system in the CBM experiment consists of a combination of absorbers and detector layers, as shown in the figure~\ref{much_detector_setup}. Due to its compact geometry, the gaps for MuCh tracking stations between the absorbers are kept as small as possible ($\sim$~10~cm gap is available between two layers of a station). 
A compact cooling system is required to meet the constraint. It is accomplished by pursuing a novel concept of using a single metal plate (with water channels grooved within) which can serve both as a mounting structure for the detector as well as a heat sink. The cooling system consists of several components i.e. a cooling plate, microcontroller board, temperature sensors, water chiller, and suction pump. An aluminium plate with water channels inside serves the purpose of heat sink.

\begin{figure}[h]
  \centering
  \includegraphics[width=7cm]{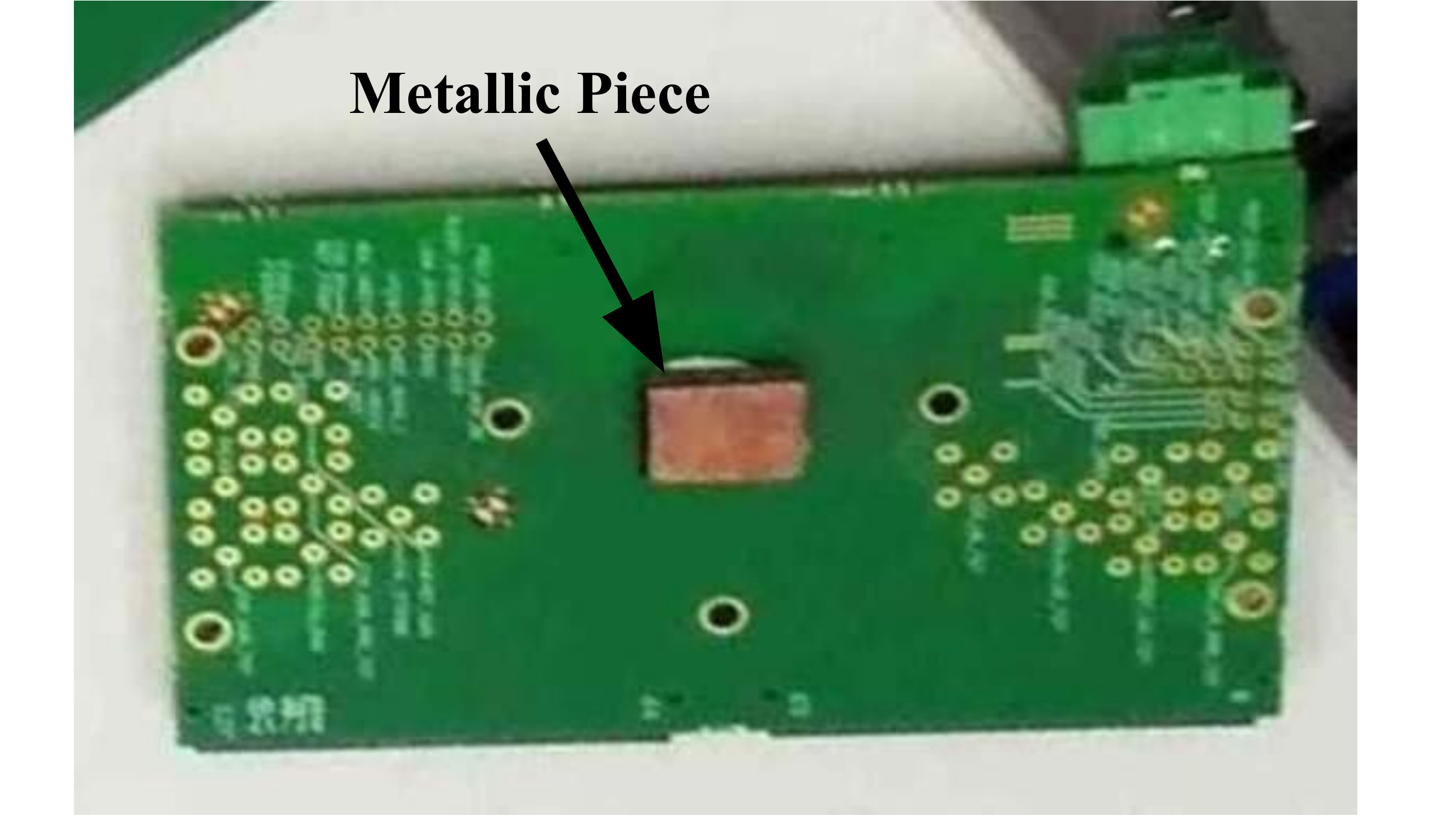}
  \caption{Back side of a FEB with metallic contact on it.}
  \label{FEB}
\end{figure}

The FEBs of the detector module with metallic contact on the bottom side of PCB as shown in figure~\ref{FEB} have been installed on one side of the cooling plate just above the water channels, using thermal glue for effective heat transmission. The heat generated from the FEBs is transmitted to the aluminium plate via this metallic contact. A submersible suction pump is used to re-circulate chilled demineralized water through the water channels within the plate to drain the heat dissipated by the FEBs. Analog LM35 temperature sensors~\cite{citeLM35} are mounted on the surface of the cooling plate to measure the temperature of the heat sink (cooling plate).

\begin{figure}[h]
  \centering
  \includegraphics[width=8cm]{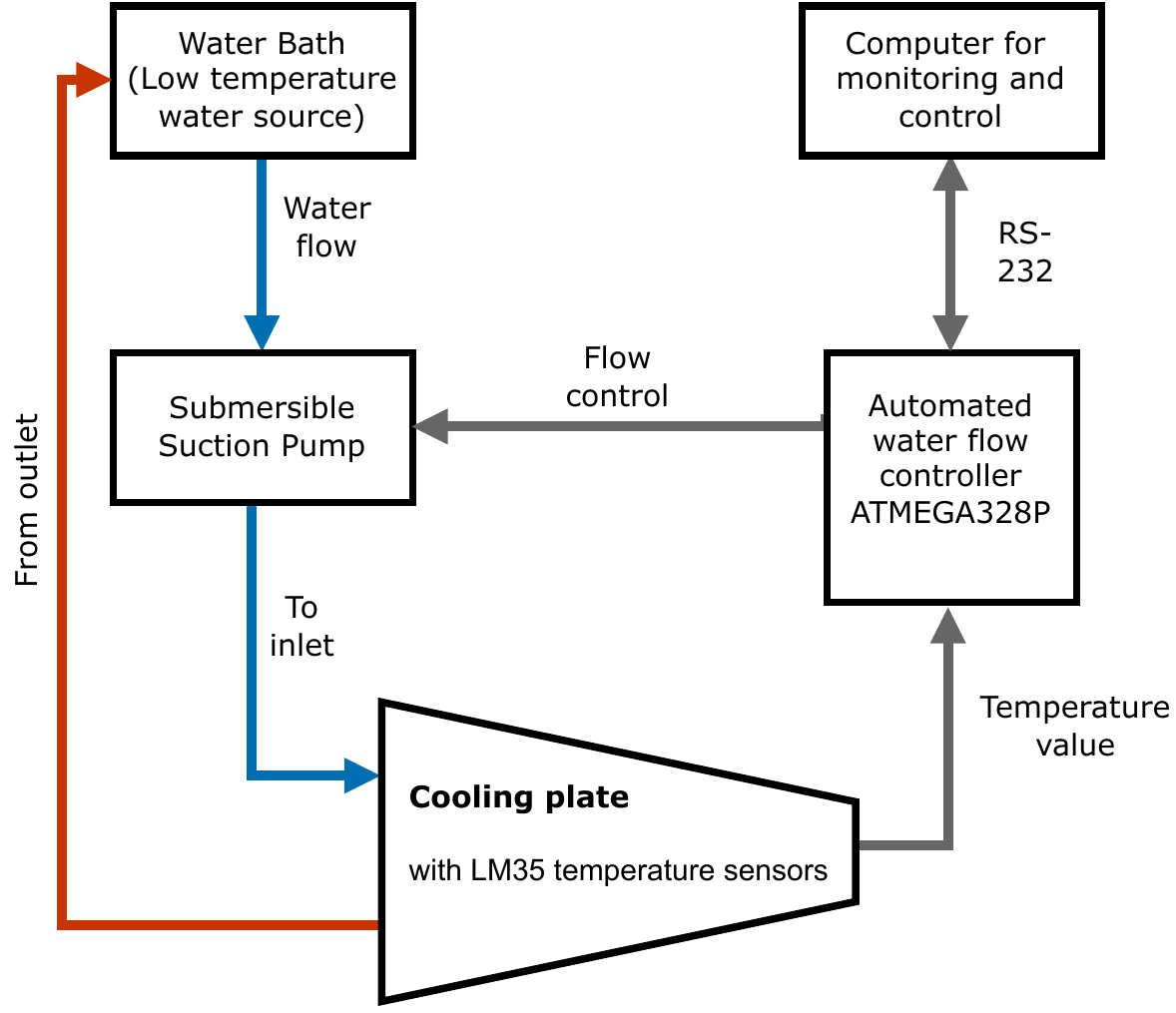}
  \caption{Flow Chart: Working principle of the cooling system.}
  \label{figFlowchartWorkingPrinciple}
\end{figure}

A microcontroller~\cite{citeMicrocontroller} based unit (see section~\ref{SecFlowControl}) is used to keep the temperature of the heat sink at a constant value. By comparing the data from the temperature sensors to a reference temperature value, the microcontroller regulates the pump speed and hence the water flow. A reference temperature is established as an external parameter in the microcontroller using a computer interface. Figure~\ref{figFlowchartWorkingPrinciple} shows the flow chart of the working principle.

When the temperature of plate rises above the reference temperature, the microcontroller turns on the re-circulation of chilled water through the plate and when the temperature of the plate falls below the reference temperature, it turns the flow off. The cycle is automated, and the temperature of the cooling plate is maintained at a predetermined reference point.   

As discussed, the use of the cooling plate is twofold: as a supporting structure for mounting the detector modules and FEBs, and as a heat sink for cooling. The dimension, design, and the type of material of the cooling plate are therefore determined considering the factors like mechanical strength, planarity, ease in building water channels inside it, and property of good heat transfer. During the experiment, the system will be operated in a high-radiation zone, hence the material for the plate should have features like low radiation activity.

\section*{III. Copper-based small size prototype}\label{SecSmallSizePrototype}
\begin{figure}[h]
  \centering 
  \includegraphics[width=6cm]{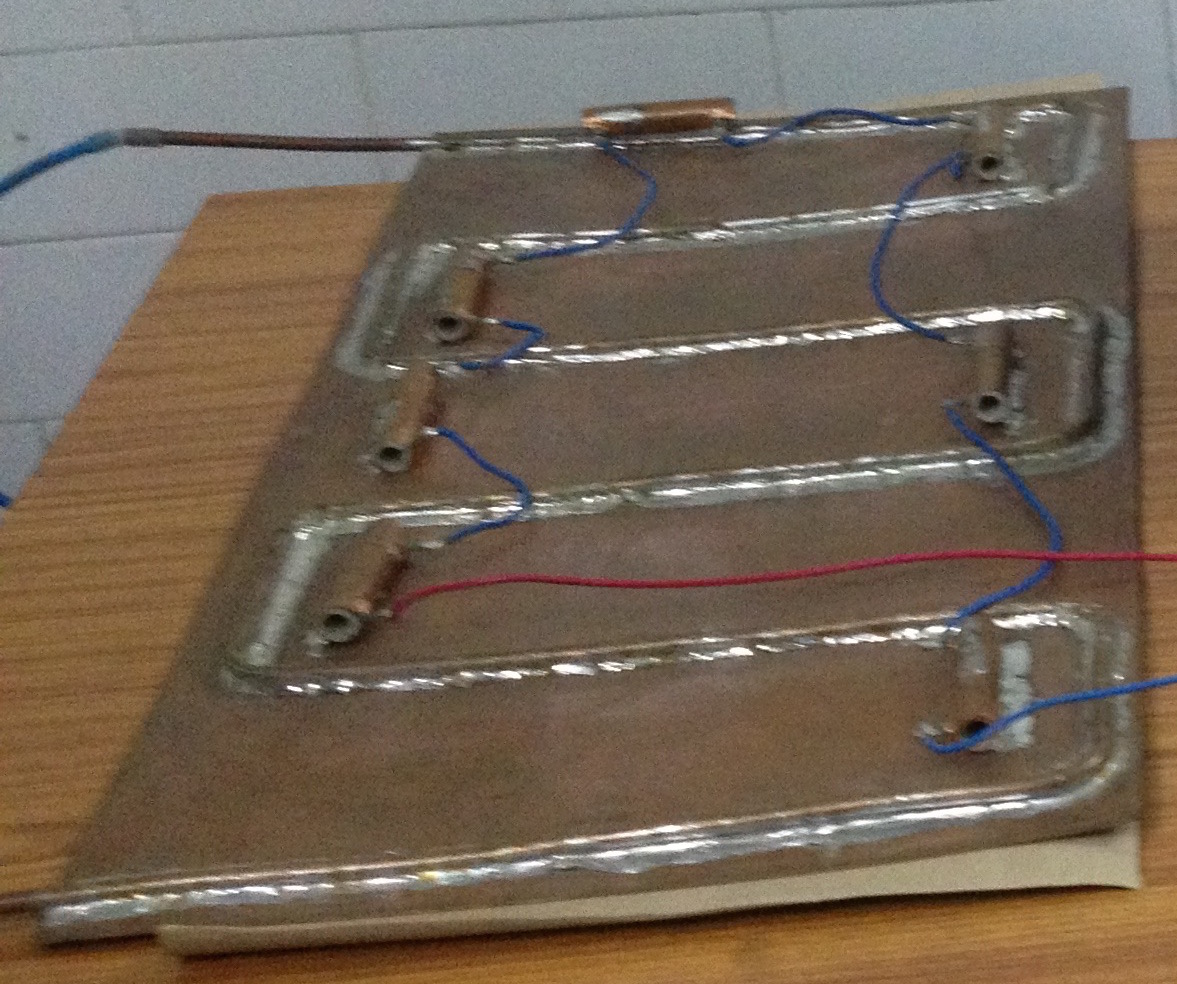}\\
  \includegraphics[width=6.4cm]{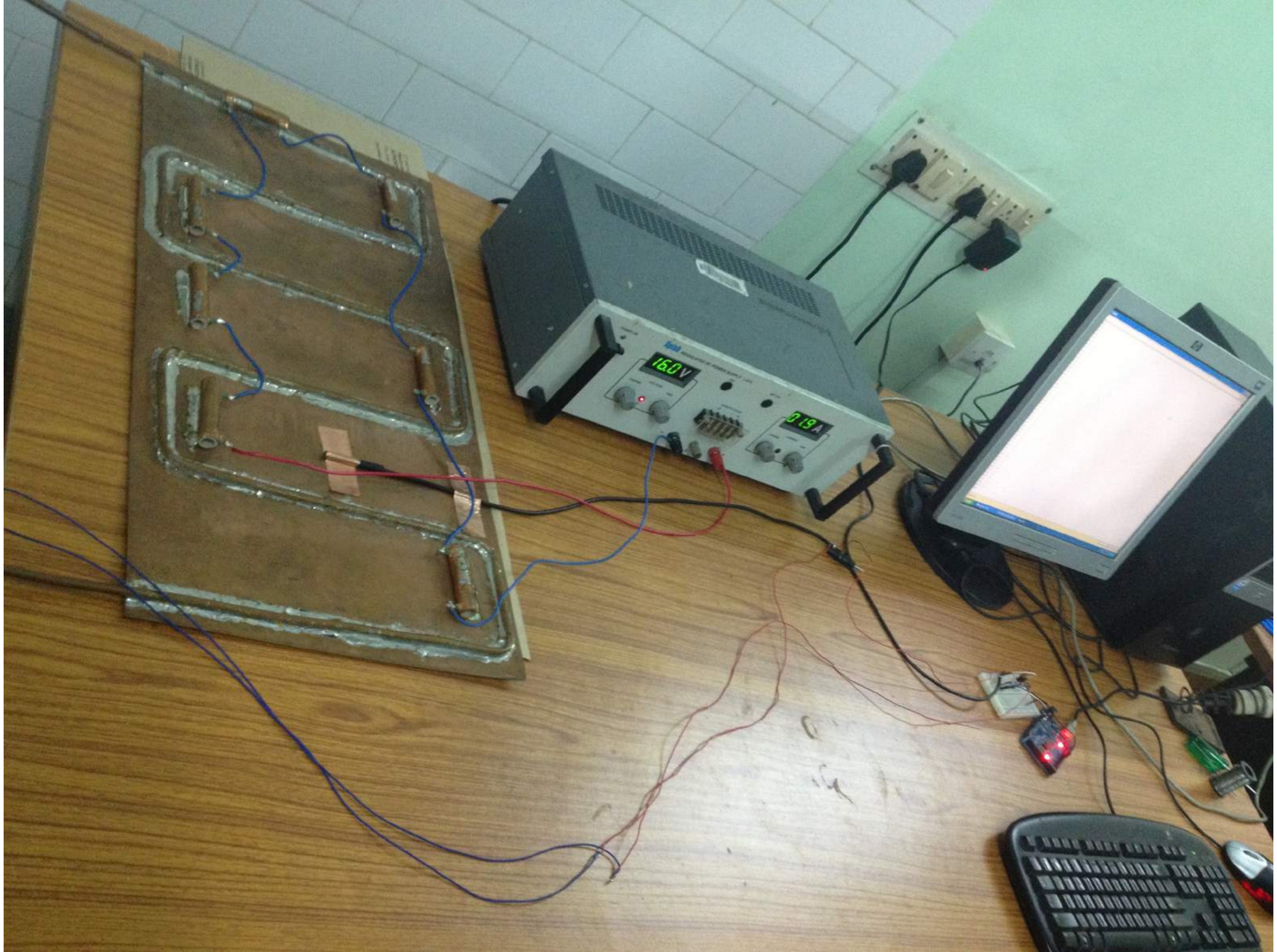}
  \caption{Small prototype: 1~mm thick copper cooling plate with copper tubes
    and heating elements brazed on it (upper), the setup of
    the cooling system (lower).}
  \label{figPrototyTest}
\end{figure}

A small-size prototype test is conducted in the laboratory to validate the proof-of-principle of the conceptual design of the cooling system~\cite{cite_Vikas_jain_dae_proceeding,cite_d_nag_dae_proceeding}. As shown in figure~\ref{figPrototyTest}~(upper), a copper plate of thickness 1~mm and size 300~mm $\times$ 600~mm with copper tubes (6~mm in diameter) brazed on its top surface is used as heat sink. 

Seven heating elements (coil resistors of 4~W dissipative power each)
are also brazed on top surface of the copper plate
(figure~\ref{figPrototyTest}~(upper)) to emulate 28~W of heating load and powered by an external regulated power supply. A test setup is prepared as shown in figure~\ref{figPrototyTest}~(lower) by appropriately connecting various components of the cooling system such as water tubes, suction pump, microcontroller, and temperature sensor etc.
\begin{figure}[h]
  \centering 
  \includegraphics[width=9cm]{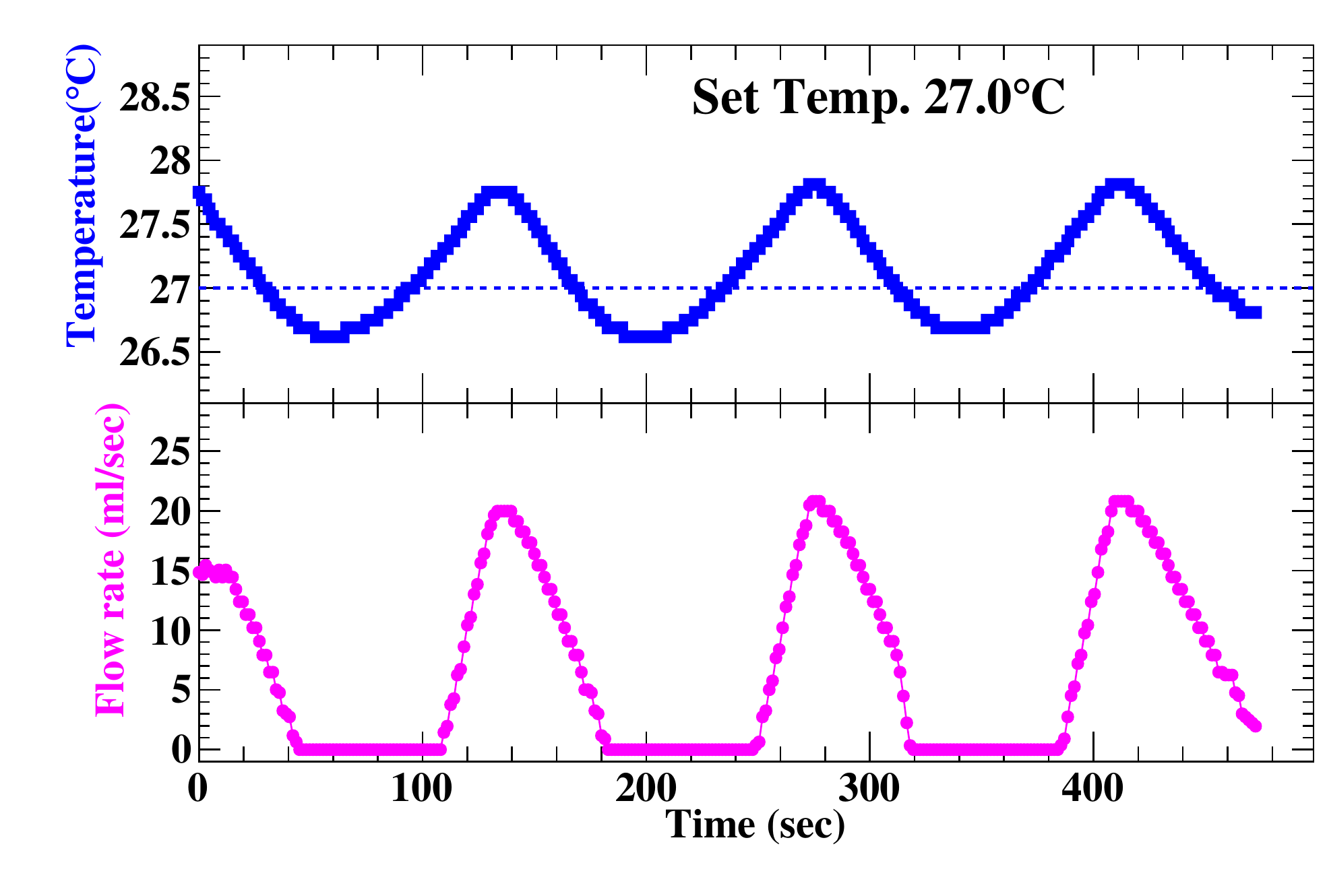}
  \caption{(Color online) Surface temperature of the copper plate and water flow rate as a 
    function of time at a reference temperature of 27$^\circ$C.}
  \label{figTempFlowVsTimeProtoType}
\end{figure}

The heating elements were turned ON to test the performance of the cooling system. The temperature of the copper plate and the water flow rate are observed for a few hours. The temperature of the copper plate went up in the beginning. As soon as it exceeded the reference temperature, re-circulation of chilled water through the copper tubes began and continued till the temperature dropped down below the reference temperature.

Figure~\ref{figTempFlowVsTimeProtoType} shows the variation of temperature and water flow rate as a function of time for a selected period. The reference temperature is set at 27$\rm {^\circ}$C for this observation. The system is clearly able to maintain the temperature of the plate at a specified value by draining off the dissipated heat. The experiment is performed for various values of the set temperature. A variation of around $\pm$0.7$^\circ$C in the temperature about the reference value is observed, which is well within the acceptable range. 

\section*{IV. Mechanical design of real-size cooling plate}\label{SecMechanicalDesign}
Considering several factors like radiation hardness, availability, and cost-effectiveness, an aluminium plate of thickness 10 --
12 mm (based on fabrication technique) is suitable to build the cooling plate. The size of the plates of first station modules is kept as 850~mm $\times$ 550~mm which is slightly bigger than the size of a MuCh module in the first station. One of the main challenging tasks in the mechanical design of the cooling plate is the fabrication of the water channels of 5~mm (6~mm) diameter inside a 10~mm (12~mm) thick aluminium plate without compromising mechanical strength and planarity of the plate. The application of the drilling technique is also ruled out due to the larger size of the plate. Two different techniques are therefore adopted for making water channels inside the plate as discussed below:\par
\begin{figure}[h]
  \centering
  \includegraphics[width=9.5cm]{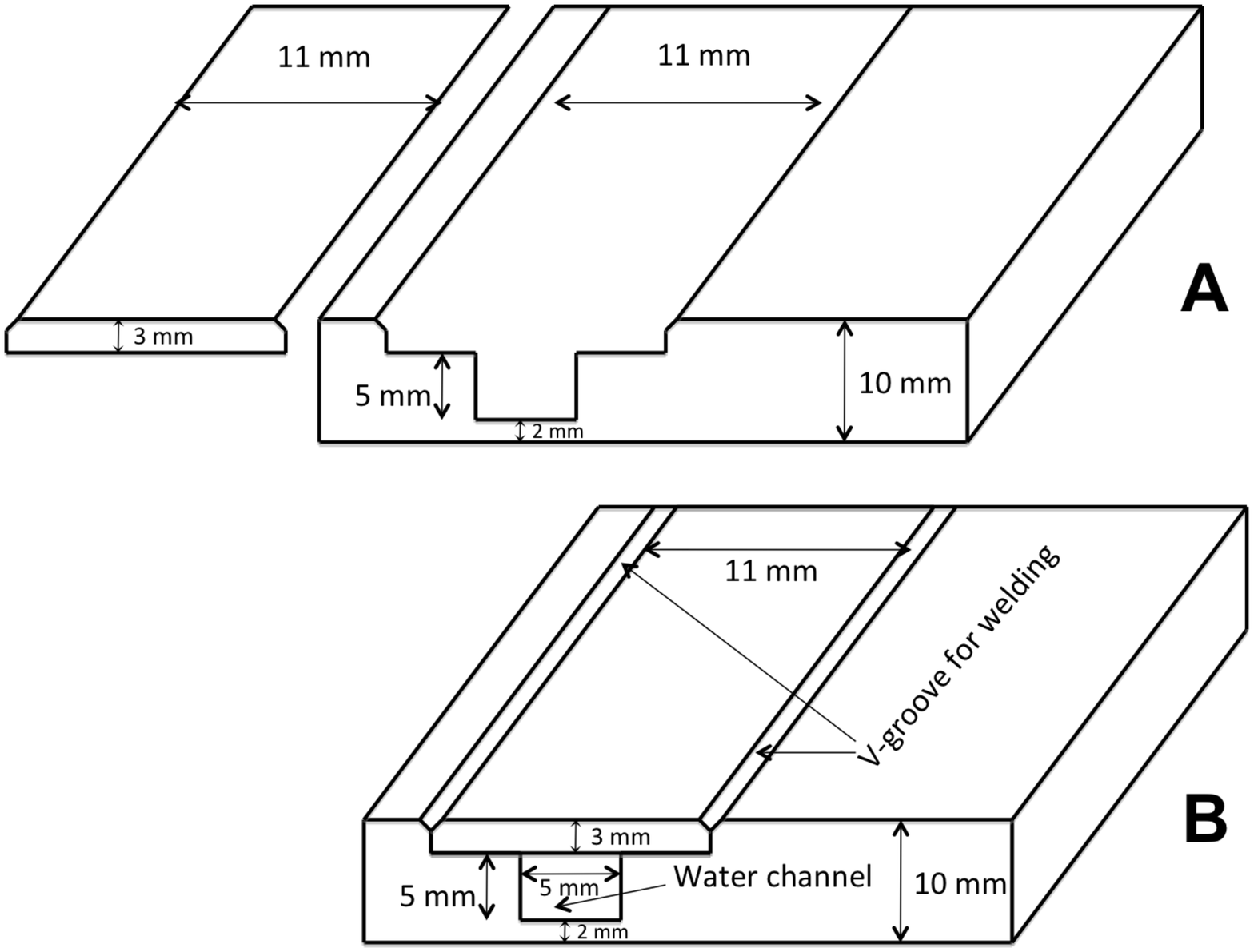}
  \caption{Building of 5~mm $\times$ 5~mm water channels inside 10~mm thick 
    aluminium plate. Preparation of T-shaped grooves (A), 
    sealed top portion of the T-shaped groove using a 3~mm 
    aluminium sheet (B).}
  \label{figTshapeGroove_00}
\end{figure}

(I) Technique--I: In the first technique, on a 10~mm thick aluminium plate, T-shaped grooves are made from the top side. First, an 11~mm wide and 3~mm thick portion is removed along the length of the plate, then a 5~mm thick and 5~mm wide groove is made, realizing a T-shaped groove as shown in figure~\ref{figTshapeGroove_00}~(A). The top portion (11~mm $\times$ 3~mm) of the groove is then sealed by welding a separate aluminium sheet of same dimension using a V-groove junction along the length of the plate as shown in figure~\ref{figTshapeGroove_00}~(B). This approach generates a 5~mm $\times$ 5~mm water channel inside the cooling plate, keeping planarity of the surface of the plate intact.
\par
\begin{figure}[h]
  \centering 
  \includegraphics[width=9cm]{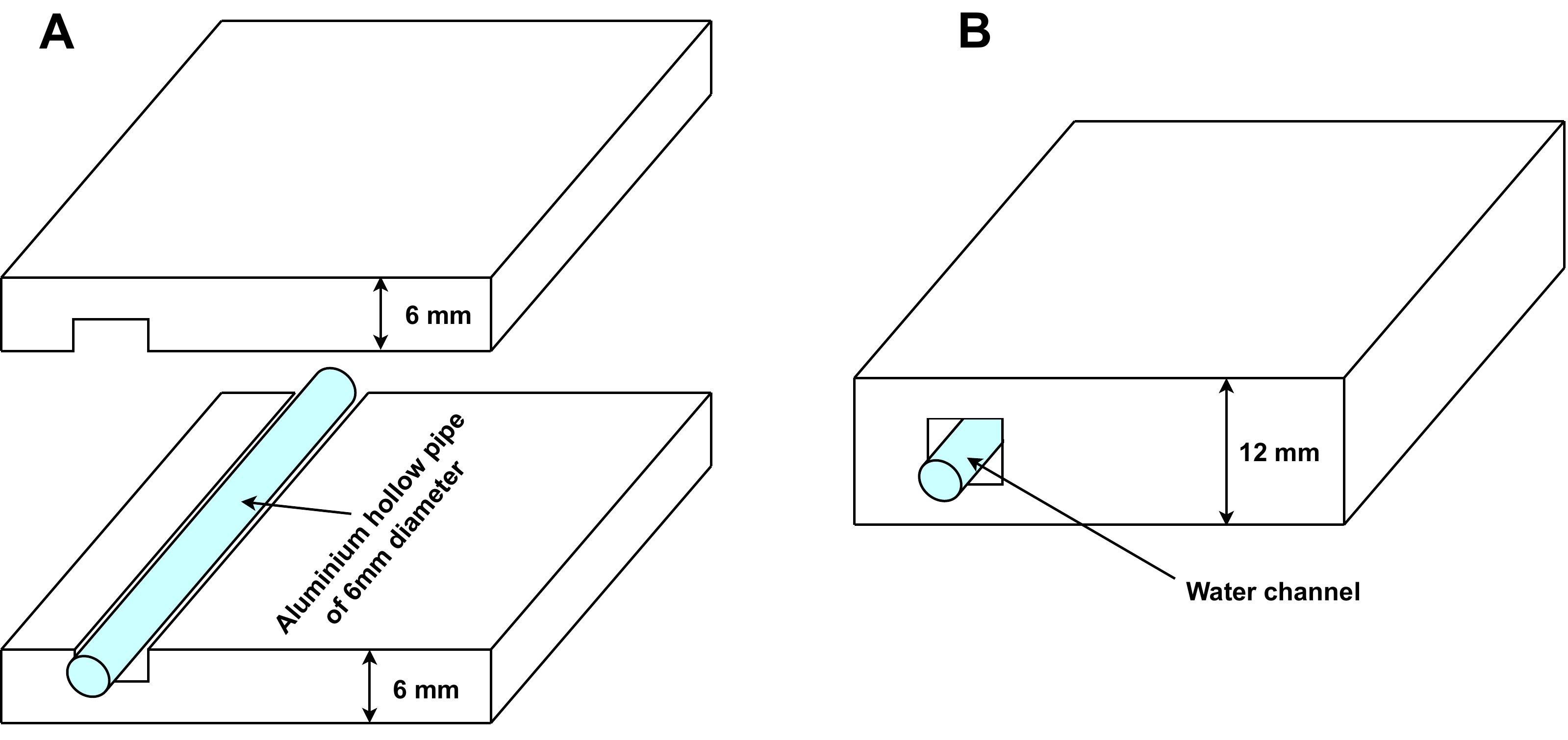}
  \caption{Building of water channels of diameter 6~mm inside 12~mm thick 
    aluminium plate. Preparation of grooves on plates in mirror image
    (A), welding two plates together   
    such that they house water pipes of diameter 6~mm inside (B).}
  \label{figGrooveScheme2}
\end{figure}

(II) Technique--II: In the second technique, two identical aluminium plates each of 6~mm thickness are taken, and grooves are made in both in mirror image as shown in figure~\ref{figGrooveScheme2}~(A). An aluminium pipe of 6~mm diameter is then press-fitted inside the groove in one of the plates and both plates are then welded together as schematically shown in figure~\ref{figGrooveScheme2}~(B) resulting in a 12~mm thick cooling plate with water channels of~6~mm diameter inside.  

\begin{figure}[h]
  \centering 
  \includegraphics[width=6cm]{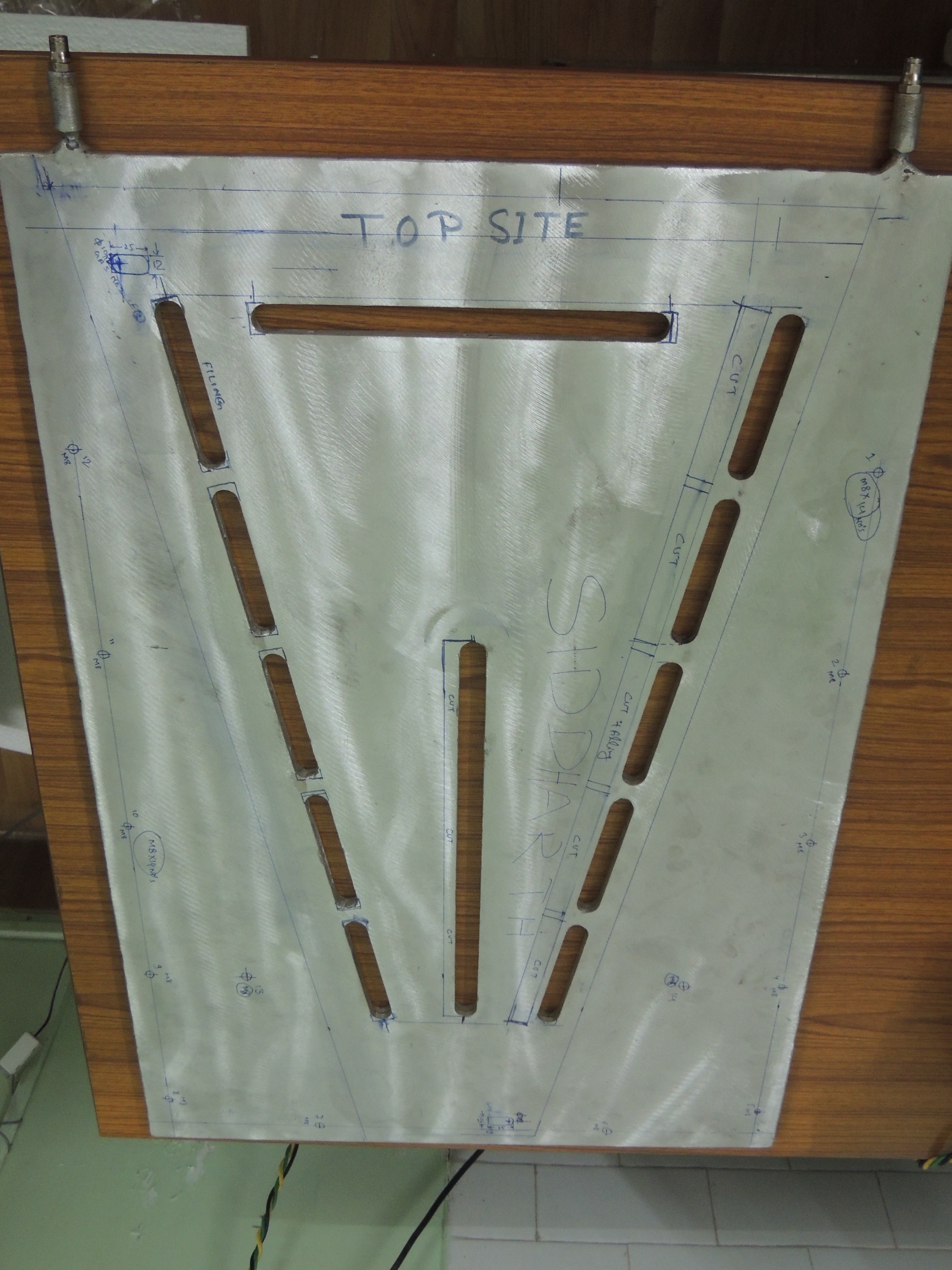}
\caption{
Sample cooling plate after fabrication using technique II.}
  \label{figAutoCad}
\end{figure}

The inlet and outlet connectors are welded from top side of the plate for connecting water tubes. Water channels are drawn in such a way that they lie precisely below the surface where FEBs are mounted to maximize the heat transfer. Figure~\ref{figAutoCad} shows one such cooling plate after the completion of the fabrication.

Cutouts are made on the plate, as shown in figure~\ref{figAutoCad} to allow connection between the detector module and FEBs using flexible cables. Both techniques have their advantages and disadvantages. Technique-I is more efficient in terms of heat transfer as the water channels are in direct contact with the cooling plate, while the Technique-II has the advantage of leak 
protection as it uses aluminium pipes. Cooling plates are built using both technique-I and technique-II and their performances are studied.

\section*{V. Control unit}\label{SecFlowControl}
\par
The control unit is one of the essential features of the cooling system because it allows keeping the surface temperature of the plate at a fixed value by regulating the coolant flow rate. The reference temperature is an external parameter and can be set or changed anytime using a computer. It also enables a facility to monitor and record temperature and coolant flow rates.

A microcontroller board ATMEGA328P~\cite{citeMicrocontroller} sits at the heart of the control unit. The board runs at 5 V external power. The controller is interfaced with a computer for monitoring and setting up reference temperature and other parameters. A negative feedback Proportional Integral Derivative (PID) algorithm runs inside the microcontroller. The input to the algorithm is the value of the temperature as measured by the temperature sensor (LM35) mounted on the surface of the cooling plate. The output of the algorithm is an 8-bit Pulse Width Modulation (PWM) signal (ranging from 0 to 255) which is used to regulate the motor speed of the suction pump. The pump is a large load, therefore the output signal is fed to the base of a power transistor which drives the suction pump. A suitable capacitor and diode in reverse bias are also connected along the terminal of the suction pump to prevent
the kickback voltage which may damage the electronics.

\section*{VI. Test setup and performances for real-size prototypes}\label{SecSetupPerformance}
To test and validate the concept, design, and integration of the cooling system with the detector modules, studies have been performed on a real-size prototype in the test beam experiment at the CERN SPS beam line facility in November 2016. Two GEM modules of the MuCh detector system were tested at the CERN SPS beamline facility using lead beam on lead target at various beam energies~\cite{cite_c_ghosh_dae_proceeding,cite_d_nag_springer_proceeding,cite_A_kumar_POP}. This experiment provided a unique opportunity to test the real size cooling prototype with an actual detector and electronics mounted on it.

\begin{figure}[h]
  \centering 
  \includegraphics[width=6cm]{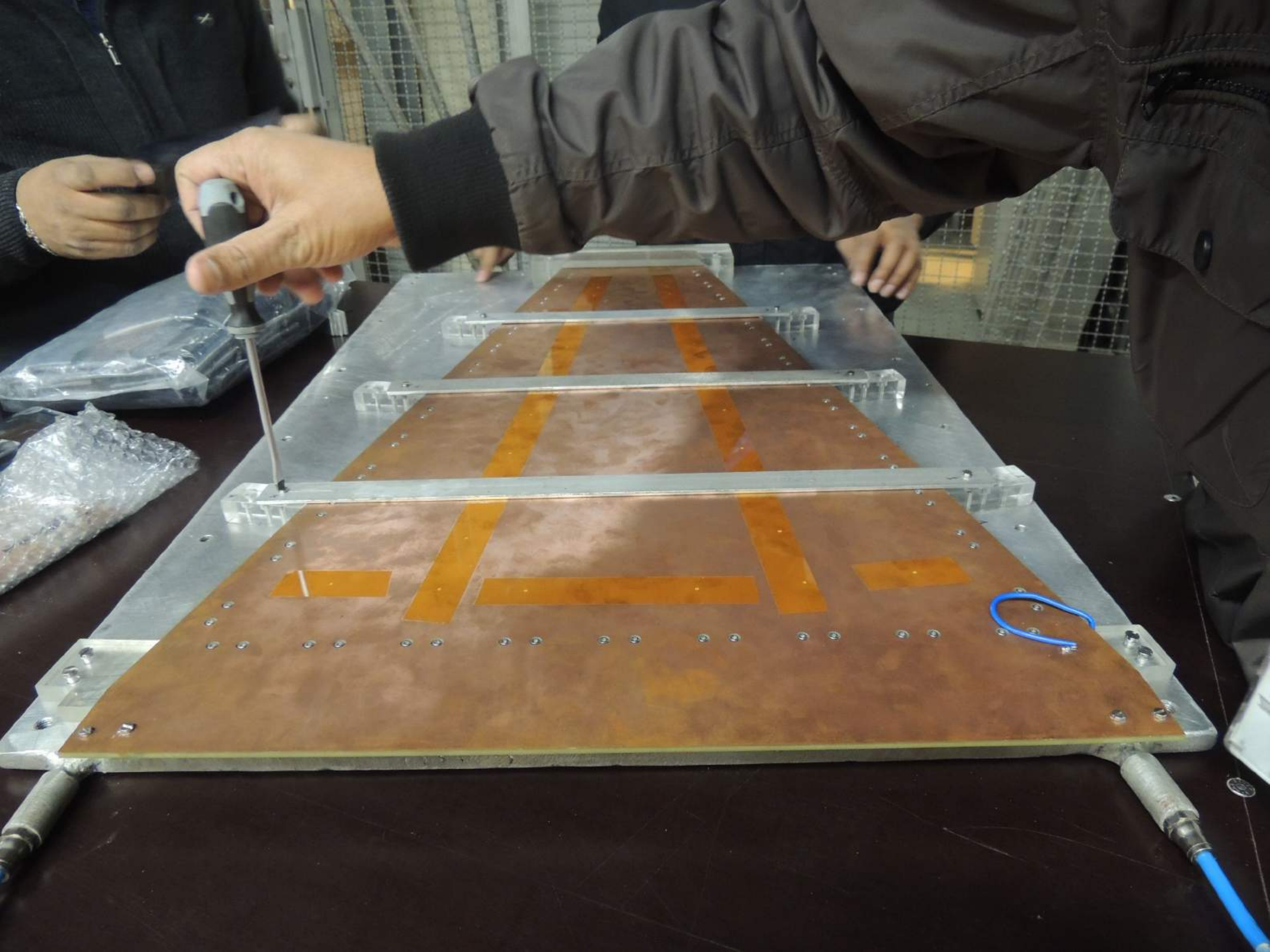}
  \includegraphics[width=6cm]{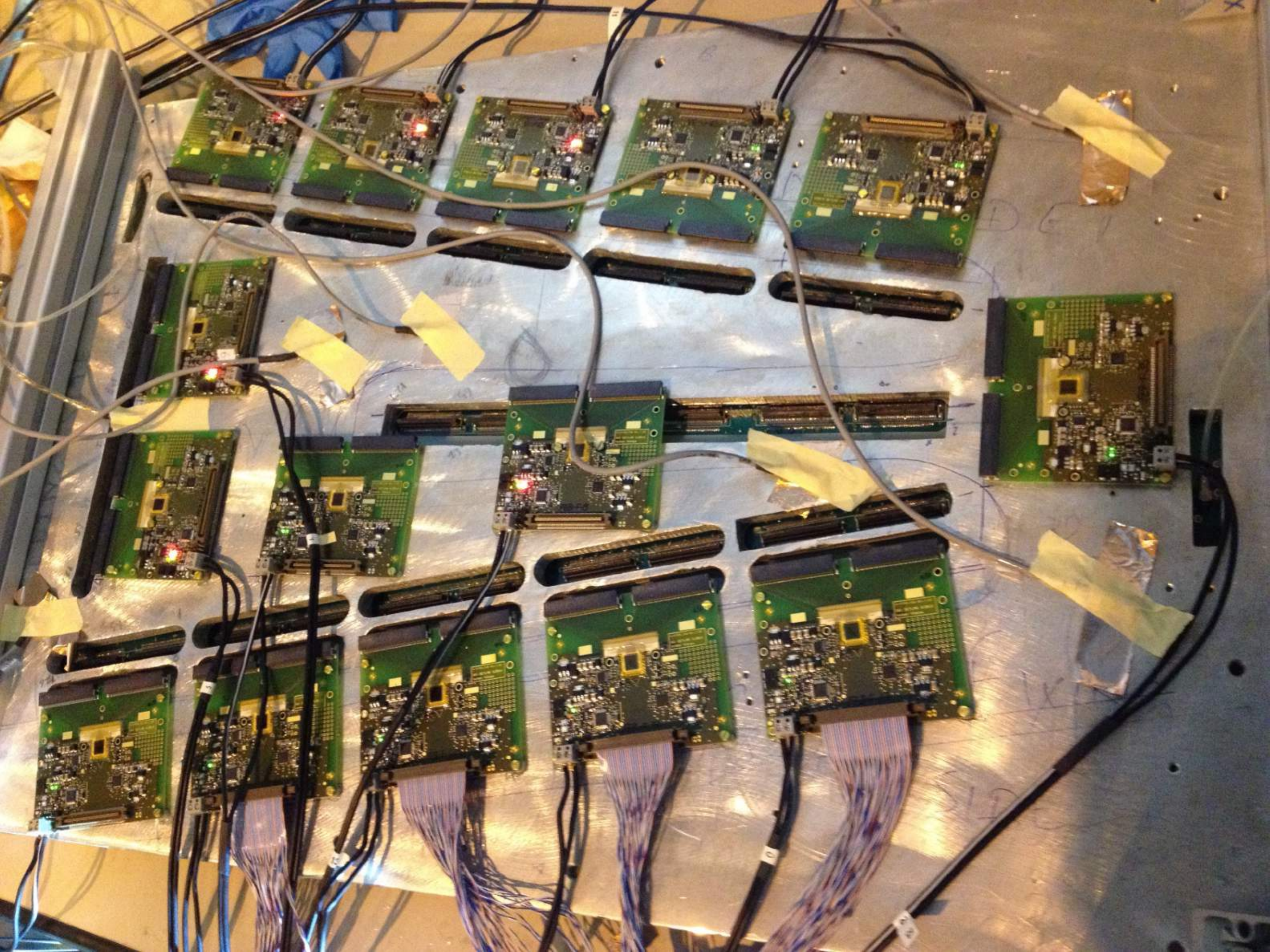}
  \caption{The mounting of the detector chamber on one side of the cooling plate (upper) and FEBs on the other side  of the same cooling plate (lower).}
  \label{figDetFebMount} 
\end{figure}

Two aluminium cooling plates were built, one following the technique--I while the other using the technique--II as discussed in section~\ref{SecMechanicalDesign}.
After preliminary tests for the water leakage, the detector chamber was mounted (see figure~\ref{figDetFebMount}~(upper)) on one side of the cooling plate while FEBs
are mounted (see figure~\ref{figDetFebMount}~(lower)) on the other side of the cooling plate using fixing screws.

\begin{figure}[h]
  \centering 
  \includegraphics[width=6cm]{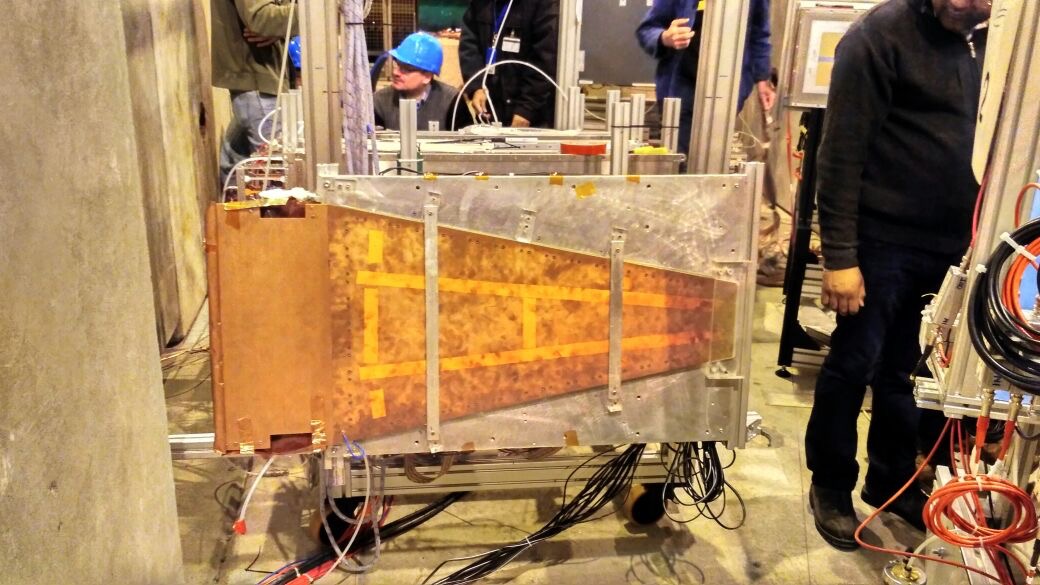}
  \includegraphics[width=6cm]{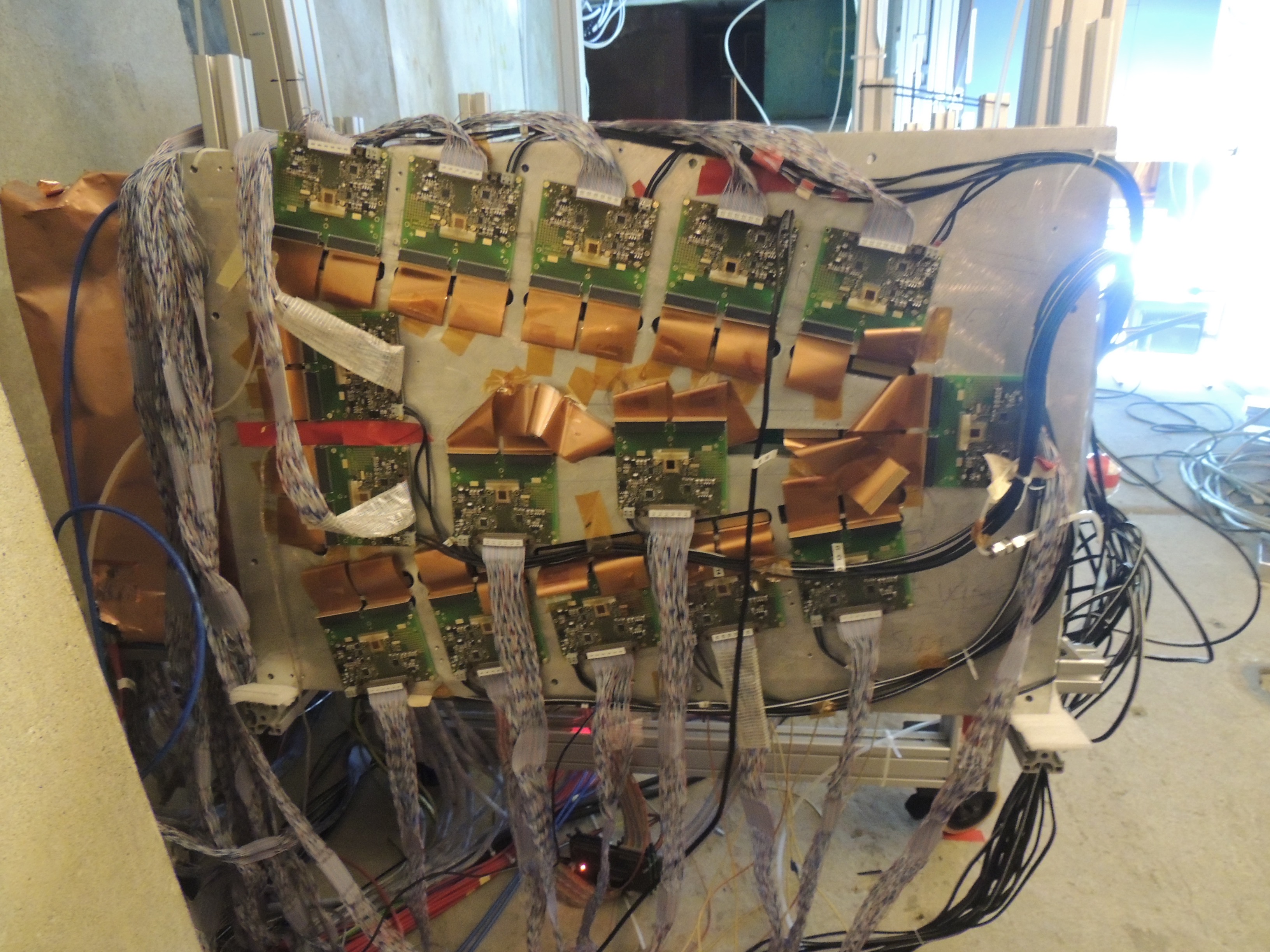}
  \caption{Integration of the cooling plate in the test beam 
    experimental setup. Upper: Detector side, Lower: FEB side.}
  \label{figInstallationTB} 
\end{figure}

A number of LM35 temperature sensors were installed at various locations on the cooling plate toward the FEB side.
Figure~\ref{figInstallationTB} shows the integration of the cooling plate in the test beam experiment from the detector (upper) and FEB (lower) sides.

\begin{figure}[h]
  \centering 
  \includegraphics[width=9.5cm]{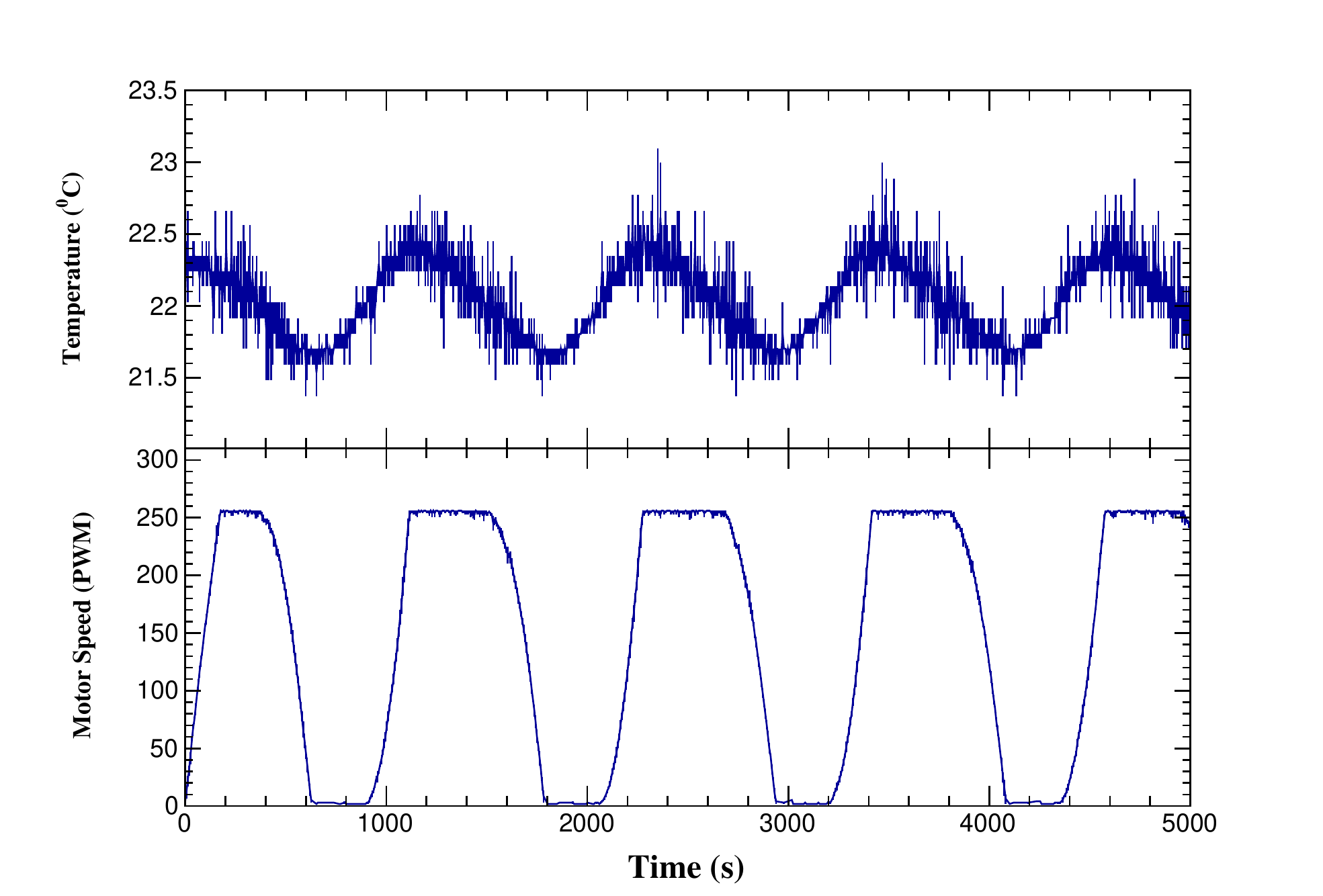}
  \caption{Surface temperature of the plate (top) and pump speed (bottom) as a function of time.}
  \label{figTempVsTimeSPSTB}
\end{figure}
Two independent water re-circulation systems were made using two submersible suction pumps and water tubes. However, a common chilled water source was used. Two separate microcontrollers were installed to regulate the speed of two pumps independently interfaced to a shared computer. The microcontrollers were kept outside the radiation zone, as they were not tested for radiation hardness. A reference temperature of 22$^{\circ}$C was set for both the systems. The chilled water was kept at a temperature of 12$^{\circ}$C. One of the temperature sensors from each plate connected to the corresponding microcontroller was used to regulate the pump speed. The data taking continued for about more than two weeks during the test beam, and FEBs remained powered-ON during this period. We observed that the cooling system for both the modules ran continuously during the entire data-taking period in an automated way without any intervention. The temperatures of the cooling plates were monitored remotely using a monitoring PC, and data points were recorded with a time stamp. The cooling performances for both the modules were quite remarkable insuring a stable temperature with $\pm$0.5$^{\circ}$C variation on surface of both plates. Figure~\ref{figTempVsTimeSPSTB} top panel shows the temperature variation, whereas the bottom panel shows motor speed in PWM units as a function of time for one of the plates with 15 FEBs. It is observed that the temperature remains constant at 22$^{\circ}$C (reference temperature) with a variation of $\pm$0.5$^{\circ}$C. It is also noticed that depending on the temperature of the plate, the pump speed varies with time. The motor speed increases when the plate temperature is higher than the reference temperature, attaining the maximum value. As soon as the plate temperature falls below the reference temperature, the motor speed slowly drops to zero, stopping the coolant flow through the plate. The cycle continues with time, keeping the temperature of the plate constant. The results are shown for a selected period; however the same behavior is obtained for more than two weeks of data taking. Similar performances are also observed for the second plate with 9 FEBs (not shown here). 

\section*{VII. Feasibility study of water distribution with multiple prototypes}\label{SecResult}
\begin{figure}[h]
  \centering 
   \includegraphics[scale=0.07,trim={18cm 0 16cm 0},clip=true]{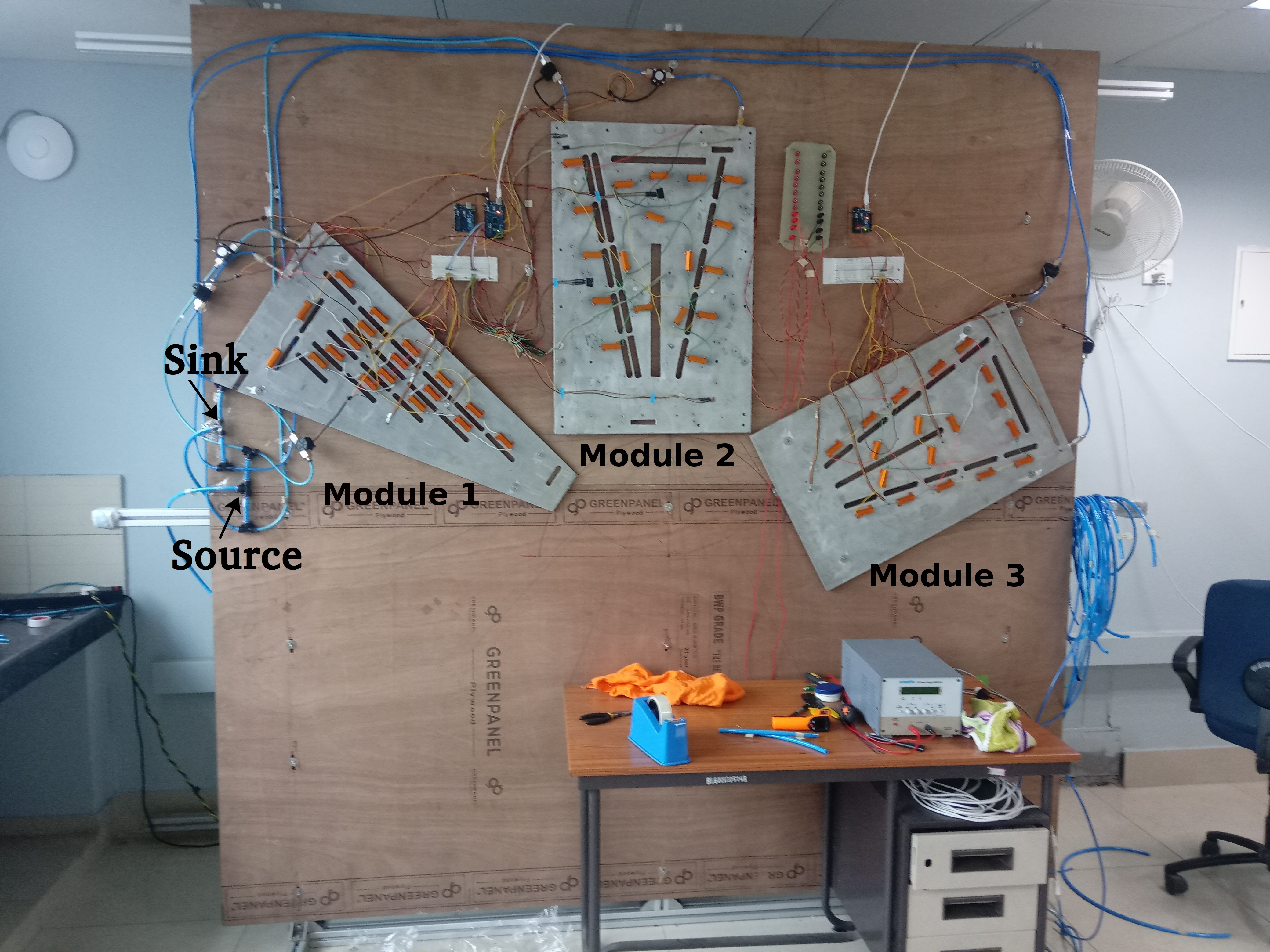}
  \caption{Test setup of three modules build at Bose Institute.}
  \label{module_setup}. 
\end{figure}

In the actual experiment, sixteen modules will be connected to each layer of the first tracking station of the MuCh detector system. A test frame made of aluminium extrusion and plywood~\cite{cite_cbm_progress_report, Kundu:2022chc} is developed at Bose Institute, Kolkata, to check how these sixteen modules should be assembled on each layer. Three modules are attached to this frame to test different arrangements or configurations, as shown in figure~\ref{module_setup}.

In absence of FEBs, the 10 $\Omega$ resistors are used as heating elements to produce uniform heating around 72~W over the whole aluminium module which is comparable to actual experimental setup. In each module, eighteen such resistors are embedded. The flat surface of each resistor is attached to the plate using thermal tape to easily transfer the heat to the modules. Heat to be dissipated is generated by passing a current through the resistors, this mechanism is called joule heating. A variable DC power supply of 30 V (Scientific DC power supply PSD3210) with a maximum current of 10 A was used to produce the joule heating. Different configurations of resistors are used to select an optimized arrangement that can generate the amount of heat expected to be produce using FEB board in the actual experimental situation. The LM35 temperature sensors are used
to measure temperature at different positions of the cooling plate. The study was carried out for two different configurations of water distribution which are as follows:

1) {\it Flow of water in series through three modules:}
In this configuration, modules are connected in series. Water from the source goes to the inlet of the first module, then the outlet of the first module is connected to the inlet of second module, and so on. Finally, water comes out of the outlet of third module to the sink.
In this case, the water temperature will be different for inlet of each module due to the absorption of heat in the previous module. 

2) {\it Flow of water in parallel through three modules:}
In this configuration, water comes directly from the source to the inlet of each module, and it flows back from each outlet to the sink. In this case, the initial temperature of the distilled water will be same for each module. Push-in pneumatic
connectors~\cite{push_in_connectors} and pneumatic polyurethane (PPU) tubes~\cite{PPU_tubes} of 10 mm and 6 mm are used in these configurations.

Six temperature sensors are attached to each module to check the temperature of module at different points on the surface. Flow rate sensors~\cite{flow_rate_sensor}  are used at the inlet and outlet of each module. A water cooler provides the cool distilled water of temperature between 16$^{\circ}$--19$^{\circ}$C. The room temperature was maintained at 25$^{\circ}$C. First, the heating starts by input voltage to the heating elements. The temperature of the modules keep rising and maximum temperature could be measured by temperature sensors. Then after some time, the water flow is switched on through the modules, due to which the temperature drops down, and it comes to a stable value. Out of the six temperature sensors, one is selected based on its location near the heating elements on the surface to compare the results. The same procedure is followed for both configurations, and results are shown in Table~\ref{table_1} and Table~\ref{table_2}.
 
\begin{table}[h!]
\centering 
  \begin{tabular}{|c|c|c|c|c|c|}
    \cline{0-5}
    & & \multicolumn{2}{|c|}{Temp.  attained ($\pm$1$^{\circ}$C)} \\
    \cline{3-6}
    &  Heat & after  & after \\ 
    Module&produced( W ) & heating ($^{\circ}$C) & cooling ($^{\circ}$C)  \\ 
    \cline{3-4}\cline{5-6}
    \hline 
    1 & 72.5  & 30 &  21 \\ 
    \hline 
    2 & 78.2 & 30 & 24 \\ 
    \hline 
    3 & 82.2  & 35 & 26 \\ 
    \hline 
  \end{tabular}
  \caption{\textit{Performance of modules when connected in series.}}
  \label{table_1}
\end{table}

\begin{table}[h!]
\centering 
  \begin{tabular}{|c|c|c|c|c|c|}
    \cline{0-5}
   & & \multicolumn{2}{|c|}{Temp.  attained ($\pm$1$^{\circ}$C)} \\
    \cline{3-6}
    &  Heat & after  & after \\ 
    Module&produced( W ) & heating ($^{\circ}$C) & cooling ($^{\circ}$C)  \\ 
    \cline{3-4}\cline{5-6}
    \hline 
    1 & 72.5 & 31 & 21 \\ 
    \hline 
    2 & 78.2 & 30 & 24 \\ 
    \hline 
    3 & 82.2 & 35 & 22 \\ 
    \hline 
  \end{tabular}
  \caption{\textit{Performance of modules when connected in parallel.}}
  \label{table_2}
\end{table}  
In both the configurations, the minimum temperature attained by the modules shows that the effect of cooling was similar for the first two modules, but it is improved for the third module. It is observed that both configurations can keep the temperature of the modules in the required range of 20$^{\circ}$--25$^{\circ}$C. However, temperature of the three modules are more uniform in parallel configurations than the series configuration. In the actual experiment, the simultaneous start of heating and cooling can avoid an initial rise in temperature. 

\section*{VIII. Summary and Outlook}\label{SecSummaryAndOutlook}
In summary, we presented details of the concept, design, fabrication, and test performances of a water-based cooling prototype under investigation for the MuCh detector system in the CBM experiment. One small size and two real-size prototypes of the cooling system were developed, and their performances were studied in detail. A setup of the three prototype modules was developed at Bose Institute, Kolkata, where two different configurations were studied. 

Tests for the small-size prototype were conducted using a copper plate and an emulated heat load in the laboratory while the real-size prototypes were tested during the test beam experiment at CERN SPS H4 beamline facility with the actual heat load and under realistic experimental conditions.

The results obtained from the study of copper plate-based small size prototype validated the proof-of-principle of the concept and design of the cooling system. The real-size prototypes were built with aluminium plates using two different mechanical techniques. A novel concept of control mechanism is realized and tested by integrating ATMEGA328P microcontroller with the cooling system to monitor and control the temperature of the cooling plate.

The experiences of operating the real-size prototypes during the test beam experiment at CERN SPS and the test results confirmed that the system under investigation is feasible in all aspects and efficient enough to meet the cooling requirement of the MuCh detector system in the CBM experiment. Three modules are also tested in both series and parallel configurations to select the suitable configuration for each layer of tracking station. It is found that parallel configuration would be a better option for water distribution of modules in the final experiment.

Several other challenges are, however, to be further investigated and addressed before realizing the full cooling setup for the integration in the experiment. Two most important issues under consideration are the scalability of the system to meet the experimental requirement and development and integration of a leak detection and protection unit or a leak less coolant flow mechanism to the existing system.

\acknowledgments
We acknowledge Mr. G. S. N. Murthy for the valuable discussions about the concept and design considerations of the cooling system. Mr. Dipanjan Nag is acknowledged for developing the microcontroller based automatic flow control system. The mechanical workshop of VECC is acknowledged for building one of the cooling plates. We thank Mr. Jayant Kumar of VECC for providing all necessary engineering drawings. Mr. Biswabibek Bandyopadhyay, Sourav Roy, and Rama Prasad Adak of Bose Institute are acknowledged for their help in developing the control unit and during the tests in the laboratory. S. K. Kundu thanks the Council of Scientific and Industrial Research (CSIR), New Delhi for financial support (File
No. 09/1022(0051)/2018-EMR-I).

\end{document}